\documentstyle[prl,aps,twocolumn,epsf]{revtex}
\begin{document}
\draft
\title{Coulomb blockade of tunneling between disordered conductors}
\author{J\"org Rollb\"uhler and Hermann Grabert}
\address{Fakult\"at f\"ur Physik, Albert--Ludwigs--Universit\"at,
 D--79104 Freiburg, Germany}
\date{Date: August 30, 2001}
\maketitle
\begin{abstract}
We determine the zero--bias anomaly of the conductance of tunnel 
junctions by an approach unifying the conventional Coulomb blockade 
theory for ultrasmall junctions with the diffusive anomalies in 
disordered conductors. Both, electron--electron interactions within the 
electrodes and electron--hole interactions between the electrodes
are taken into account non--perturbatively. Explicit results
are given for one-- and two--dimensional junctions and the 
crossover to ultrasmall junctions is discussed.
\end{abstract}

\vspace{-0.5cm}

Coulomb blockade of tunneling 
\cite{Nato92} has been
investigated extensively in the last decade. Most of these studies
were focussed on ultrasmall tunnel junctions with small capacitance $C$.
Then, the relevant energy scale is the charging energy 
$E_{C} = e^{2}/2C$. In these effectively zero--dimensional junctions the
charge of a tunneling electron is distributed over the junction area
fast compared to the time scale $\hbar/eV$ associated with
the applied voltage $V$, and Coulomb blockade
effects are observable for single junctions only because of the slow
charge relaxation via the environmental impedance $Z(\omega)$
\cite{Devoret90andGirvin90}.

The situation changes for tunnel junctions with large lateral
extensions where the time scale of propagation of the electromagnetic
field over the system size becomes relevant.
Then screening and relaxation processes within the electrodes need to be
considered explicitly. Particularly, in junctions with disordered
low--dimensional electrodes, the field propagates diffusively, and a
suppression of the tunneling conductance arises from the reduction 
of the electron density of states by electron--electron interactions 
\cite{AA79,AAL80}.

The pioneering papers on the diffusive anomalies in disordered 
conductors 
\cite{AA79,AAL80} 
are based on perturbation theory 
in the Coulomb interaction and fail at very low energy scales.
As is well--known from the conventional theory of Coulomb blockade
in effectively zero--dimensional junctions 
\cite{Nato92,Devoret90andGirvin90}
a non--perturbative approach is needed for a complete 
treatment of the Zero--Bias Anomaly (ZBA) of the conductance.
Furthermore, apart from
the interaction within the electrodes, the Coulomb interaction 
between the electrodes needs to be taken into account.
Only in limiting cases, e.g., for a strongly disordered
electrode screened by a bulky electrode, can the mutual
interaction be incorporated into a screened Coulomb potential.

Here, we present a microscopic approach 
treating all low energy excitations of the Coulomb field 
non--perturbatively and obtain explicit results for 
the current--voltage relation of zero--, one--, and two--dimensional
junctions at finite temperature.
Specifically, we consider a tunnel junction consisting of two
disordered metallic electrodes separated by a thin insulating barrier 
with a chemical potential difference $\mu_{1} - \mu_{2} = eV$.
The tunneling current $I$ can be calculated from the tunneling 
Hamiltonian 
\vspace{-0.1cm}
\begin{eqnarray}
 H_{T} = \sum_{k,q} T_{kq} c_{k}^{\dagger} c_{q}^{} 
  + \rm{h.c.} ,
\end{eqnarray}
where  $k$ and $q$ label the momenta of electrons in the right and
left electrode, respectively \cite{remark_spin}. 
To leading order in $H_{T}$ the current reads
\begin{eqnarray}\label{current_general}
 I(V) = - 2 e \; {\rm Im}[X^{\rm ret}(eV)].
\end{eqnarray}
Here $X^{\rm ret}(\omega)$ is the Fourier transform of 
\begin{eqnarray}\label{fourpointfunction}
 X^{\rm ret}(t) = -i \theta(t) 
  \sum_{k,q} \sum_{k',q'} T^{}_{kq} T^{*}_{k'q'} 
  \langle  
   [c^{\dagger}_{k}(t) c^{}_{q}(t), c^{\dagger}_{q'} c^{}_{k'}] 
  \rangle 
\end{eqnarray}
where we put $\hbar=1$, 
and the time evolution of the Fermi operators arises from the
Hamiltonian in the absence of tunneling, but in the presence of
disorder and interactions.

To evaluate the expectation value with four Fermi operators
in Eq.~(\ref{fourpointfunction}), we employ the 
Keldysh $\sigma$--model technique \cite{Belitz94,Kamenev99}
leading to the result in Eq.~(\ref{current_final}) below.
We start from the partition function
\begin{eqnarray}
 Z = {\cal N} \int {\cal D} \bar{\psi}_{1} {\cal D} \psi_{1} 
  {\cal D} \bar{\psi}_{2} {\cal D} \psi_{2} {\cal D} \phi \; 
  e^{i S[\bar{\psi}_{1}, \psi_{1}, \bar{\psi}_{2}, \psi_{2}, \phi]}
  \; ,
\end{eqnarray}
where $\cal{N}$ is a normalization constant. The action is given by
\vspace{-0.1cm}
\begin{eqnarray}
 S[\bar{\psi}_{j}, \psi_{j}, \phi]
 &=& \sum_{j} \int_{\cal C} dt \int_{V_{j}} dr \; 
  \bar{\psi}_{j}(r,t) \biggl[i \frac{\partial}{\partial t} 
  +\frac{1}{2 m} \Delta \nonumber \\
 && - W_{j}(r) + e \phi(r,t)\biggr] \psi_{j}(r,t)
  \nonumber \\ 
 && + \frac{1}{8 \pi} \int_{\cal C} dt \int dr \; 
  \biggl[\nabla \phi(r,t)\biggr]^{2} ,
\end{eqnarray}
where ${\cal C}$ is the Keldysh contour, $W_{j}(r)$ are Gaussian 
distributed disorder potentials, 
$\phi$ is the electric potential, and $e$ is the electric charge. 
While the fermionic terms are integrated over the volume of the 
electrodes only, the last term is integrated over entire space,
which leads to electrostatic coupling between the electrodes. For 
simplicity we disregard polarization effects in the tunnel barrier,
that may easily be incorporated in the capacitance $C_{0}$ introduced
below. 

Assuming that the impurity potentials in the two electrodes are 
statistically independent, the Gaussian integrations over the 
realizations of $W_{j}(r)$ lead to quartic interactions
of the fermion fields within each electrode. Within the 
$\sigma$--model approach these interactions
can be decoupled by introducing Hubbard--Stratonovich matrix fields
$\hat{Q}_{j}$ \cite{Belitz94}. Subsequently, the fermionic fields can be
integrated out in the usual way yielding a representation of the 
impurity averaged partition function as a functional integral over 
the electric potential $\phi$ and the matrix fields $\hat{Q}_{j}$. 
Since for given field $\phi$ all of these transformations can be done 
independently for each electrode, the resulting action 
$S[\phi,\hat{Q}_{1},\hat{Q}_{2}]$ can immediately be inferred from 
earlier work \cite{Belitz94}.

To proceed we follow Kamenev and Andreev \cite{Kamenev99}
and evaluate the integrals over the matrix fields $\hat{Q}_{j}$ 
using the saddle point approximation. 
The saddle point solution can be obtained 
analytically for spatially uniform fields $\phi(t)$ only. Provided
$\phi$ is essentially constant on scales of the size of the mean free
path $l$, an effective action was derived in
\cite{Kamenev99} which incorporates dynamic screening in the random
phase approximation (RPA) and the diffuson vertex correction within each 
electrode. 

One is left with a Gaussian integral over the Coulomb field with a
purely electromagnetic action which determines the field
propagators. Since these depend on the geometry of the electrodes, 
we first restrict ourselves to the case of thin films. Then, for low
energy excitations with wave vectors $q \ll a^{-1}$, where $a$ is the
thickness of the films, we have effectively a two--dimensional problem,
where the Coulomb field depends on two spacial coordinates in each 
electrode (see also the discussion below). 
Although only the large distance
behavior of the Coulomb interaction matters, we have to keep the
barrier thickness $\Delta$ finite since dipolar interactions 
arise \cite{remark_delta}.
With the sources restricted to two--dimensional films at 
$z_{j} = \pm\Delta/2$, the bare Coulomb interactions reads 
\begin{eqnarray}
 U_{ij}^{(0)}(x-x',y-y') 
  = \int dz dz' \; 
  \frac{e^{2} \delta(z-z_{i}) \delta(z'-z_{j})}{|r-r'|} .
 \nonumber 
\end{eqnarray}
The resulting equation for the matrix 
$U^{\rm ret}(q,\omega)$ of the Fourier transformed field propagator
then takes the form of the diffusive RPA
\begin{eqnarray}\label{rpa_dyson}
 U^{\rm ret}(q,\omega) = \left[
  \left(U^{(0)}(q,\omega) \right)^{-1} 
  + P(q,\omega) \right]^{-1} ,
\end{eqnarray}
where
\begin{equation}\label{bare_coulomb_propagator}
 U^{(0)}(q,\omega) = \left(
  \begin{array}{cr}
  u(q) &  v(q) \\
  v(q) &  u(q) \\
  \end{array}
  \right).
\end{equation}
Here $u(q)$ and $v(q)$ are the two--dimensional bare Coulomb 
interactions inside and between the electrodes, respectively. 
Neglecting polarization effects in the barrier, we obtain 
$u(q) = 2 \pi e^{2}/q$ and
$v(q) = 2 \pi e^{2} e^{-q \Delta}/q$.
The polarization function
\begin{equation}\label{polarization_function}
 P(q,\omega) = \left(
  \begin{array}{cc}
  \nu_{1} \frac{D_{1} q^{2}}{D_{1} q^{2} - i \omega}  &  0 \\
  0  &  \nu_{2} \frac{D_{2} q^{2}}{D_{2} q^{2} - i \omega} \\
  \end{array}
  \right)
\end{equation}
contains the bare electron densities $\nu_{j}$ and the electron 
diffusion constants $D_{j}$ of the two electrodes.

Supplementing the partition function $Z$ with source fields
that couple to the fermion operators of each electrode, 
it may be employed as a generating functional to evaluate 
impurity averaged correlation functions like 
\begin{eqnarray}
 X^{>}(t) = -i \sum_{k,q} \sum_{k',q'} T^{}_{kq} T^{*}_{k'q'}
  \overline{\langle c^{\dagger}_{k}(t) c^{}_{q}(t) 
  c^{\dagger}_{q'} c^{}_{k'} \rangle}.
\end{eqnarray}
The pre--exponential factors resulting from
functional derivatives with respect to the sources
are replaced by their saddle point 
approximation. 
In terms of an average tunnel matrix element $T$, we find
\begin{eqnarray}\label{x_greater}
 X^{>}(t) &=& -i \nu_{1} \nu_{2} |T|^{2} 
  \int d\epsilon \int d\epsilon' \; e^{i(\epsilon-\epsilon') t} 
  \nonumber \\
 && \times n(\epsilon) [1-n(\epsilon')] e^{J(t)} ,
\end{eqnarray}
where the interaction effects are incorporated in 
\begin{eqnarray}\label{jt}
 J(t) = 2 \int \frac{d\omega}{2 \pi} \; 
  \frac{e^{-i \omega t}-1}{1-e^{-\beta \omega}} \; 
  {\rm Im} \, Y(\omega).
\end{eqnarray} 
Here $\beta$ is the inverse temperature and 
\begin{eqnarray}\label{c_omega_general}
 Y(\omega) = \sum_{i,j} \sum_{q} (2 \delta_{ij}-1) 
  \frac{U_{ij}(q,\omega)}
  {(D_{i} q^{2} - i \omega)(D_{j} q^{2} - i \omega)} .
\end{eqnarray} 
The factors $(D_{j} q^{2} - i \omega)^{-1}$ come from the
vertex correction terms in the action and are seen to obey the
classical diffusion equation. When the two--dimensional result for
$Y(\omega)$ is inserted into Eq.~(\ref{jt}), 
the $\omega$--integral needs
to be cut off at frequencies of order 
$|\omega| \approx 1/\tau_{0} \approx a^{2}/D$ where the crossover from
two-- to three--dimensional behavior occurs.

Now, combining Eqs.~(\ref{current_general}) and (\ref{x_greater}), 
the expression for the current is found to read
\begin{eqnarray}\label{current_final}
 I(V) = \frac{G_{0}}{e}
  \int_{-\infty}^{\infty} d\epsilon \; \epsilon P(eV - \epsilon) 
  \frac{1-e^{-\beta eV}}{1-e^{-\beta \epsilon}} ,
\end{eqnarray}
where $G_{0} = 4 \pi e^{2} \nu_{1} \nu_{2} |T|^{2}$ is the bare
tunneling conductance, and where  
we have introduced the spectral density
\begin{eqnarray}\label{p_of_e}
 P(\epsilon) = \frac{1}{\pi} {\rm Re} 
  \int _{0}^{\infty} dt \; e^{i \epsilon t} e^{J(t)}.
\end{eqnarray}
The expression (\ref{current_final}) constitutes the central 
result of this work. It is formally identical to the
conventional expression for the current--voltage relation of ultrasmall
tunnel junctions (zero--dimensional case) 
\cite{Devoret90andGirvin90}, 
and determines the non--perturbative effect of Coulomb interactions 
on the $I$--$V$--curve of spatially extended disordered tunnel 
junctions. The very same
form can also be derived in the case of effectively one--dimensional
wires (see below). 

To make contact with previous work, we first consider the (somewhat
unrealistic) case where Coulomb interactions between the electrodes 
are disregarded, i.e., $U_{12}=U_{21}=0$. Then 
$J(t) = J_{1}(t) + J_{2}(t)$ where $J_{j}(t)$ contains the
contribution from $U_{jj}$ only. Replacing $J(t)$ by
$J_{j}(t)$ in Eq.~(\ref{p_of_e}), we may define spectral densities
$P_{j}(\epsilon)$ for each electrode.
The current
(\ref{current_final}) can then be written in the familiar form
\begin{eqnarray}\label{dos_current_general}
  I &=& 4 \pi e |T|^{2} \int d\epsilon \;
   \nu_{1}(\epsilon) \nu_{2}(\epsilon - eV)
   \left[n(\epsilon - eV) - n(\epsilon) \right] 
\end{eqnarray}
with the densities of states
\begin{eqnarray}\label{dos_independent_general}
 \nu_{j}(\epsilon) = \nu_{j} \int_{-\infty}^{\infty} d\epsilon' \; 
  \frac{1+e^{-\beta \epsilon}}{1+e^{-\beta \epsilon'}}
  P_{j}(\epsilon - \epsilon').
\end{eqnarray}
This relation gives a non--perturbative result for the effective 
tunneling density of states of an electrode in presence of Coulomb 
interactions expressed in terms of the spectral density 
$P_{j}(\epsilon)$ familiar from Coulomb blockade theory. 
When $P_{j}(\epsilon)$ is replaced by its 
perturbative approximation, the seminal result by 
Altshuler, Aronov, and Lee 
\cite{AAL80} is recovered. 

In particular, for a two--dimensional film at zero temperature
the density of states (\ref{dos_independent_general}) reads
%
\begin{eqnarray}\label{dos_independent_2D}
 \nu(\epsilon) = \nu
  \exp\left[-\frac{1}{4 \pi g} \ln|\epsilon \tau_{0}| 
  \ln\left(\frac{|\epsilon|}{(D \kappa^{2})^{2} \tau_{0}} 
  \right) \right],
\end{eqnarray}
where we have suppressed the index $j$. Further,
$g = 2 \pi \nu D$ is the conductance in 
units of $e^{2}/2 \pi$ and $\kappa = 2 \pi e^{2} \nu$ 
is the inverse screening length in two dimensions. The
non--perturbative result (\ref{dos_independent_2D}) has been obtained
previously by Kamenev and Andreev \cite{Kamenev99}.

We now return to the full problem and determine the non--perturbative 
$I$--$V$--curve for two--dimensional interacting electrodes at zero 
temperature.
Then, the four terms in Eq.~(\ref{c_omega_general}) cannot be split
into contributions of each electrode, and the spectral function 
$P(\epsilon)$ characterizes the coupled system.
In the parameter range relevant for tunneling experiments 
$u(q)-v(q) \approx 2 \pi e^{2} \Delta$, and we obtain from 
Eq.~(\ref{c_omega_general})
%
\begin{eqnarray}\label{c_omega_c0}
 Y(\omega) = \frac{e^{2}}{C_{0}} 
  \sum_{q} \sum_{j=1}^{2} \frac{\lambda_{j}}
  {(D^{*} q^{2} - i \omega) (D_{j} q^{2} - i \omega)}.
\end{eqnarray}
Here we have introduced the field diffusion constant
$D^{*} = (2 \delta_{1} \delta_{2} + \delta_{1} D_{2} 
+ \delta_{2} D_{1})/(\delta_{1} + \delta_{2})$, 
with $\delta_{j} = D_{j} \kappa_{j} \Delta$ 
and the numerical factors
$\lambda_{j} = (2 \delta_{i} + D_{i} - D_{j})/
2(\delta_{1} + \delta_{2})$ ($i \ne j$). 
While $D$ and $\kappa$ are properties of the electrodes, $\Delta$ may
also be expressed in terms of the capacitance per unit area
$C_{0}=1/4 \pi \Delta$ of the junction. 
With the result (\ref{c_omega_c0}) we readily 
obtain from Eq.~(\ref{current_final}) for the zero temperature 
differential conductance $G(V)=\partial I/ \partial V$
at voltages $V \ll V_{0}$
\vspace{-0.2cm}
\begin{eqnarray}\label{general_2d}
 \frac{G(V)}{G_{0}} &=& \frac{e^{-2 \gamma/g}}{\Gamma(1+2/g)} 
  \left(\frac{V}{V_{0}}\right)^{2/g} ,
\end{eqnarray}
where $\gamma = 0.577\ldots$ is Euler's constant, 
$g = \frac{8 \pi^{2}}{e^{2}} C_{0} D^{*} \left(\sum_{j} 
\lambda_{j} \ln(1/\xi_{j})/(1-\xi_{j}) \right)^{-1}$ with
$\xi_{j} = D_{j}/D^{*}$ is a dimensionless parameter, and
$V_{0} = 2 \pi/e \tau_{0} \approx 2 \pi D/e a^{2}$.

When $g \gg 1$ we recover the perturbative result 
%
\begin{eqnarray}\label{g_perturbative_2d}
 G(V) = G_{0} [1 + 2/g \ln(V/V_{0})] ,
\end{eqnarray}
Hence in the non--perturbative approach the logarithmic corrections of 
perturbation theory are exponentiated to a power--law dependence on $V$.
In the limiting case where one of the electrodes is bulky, i.e.
$\kappa_{1} \gg \kappa_{2}$, this result can be rewritten by means
of Eq.~(\ref{dos_independent_general}) in terms of an effective
tunneling density of states of the other electrode
%
\begin{eqnarray}
 \nu(\epsilon) = \nu_{0} \, \frac{e^{-2 \gamma/g}}{\Gamma(1+2/g)} 
  \left(\frac{\epsilon}{e V_{0}} \right)^{2/g}
\end{eqnarray}
This gives
a non-perturbative generalization of the result obtained by 
Altshuler, Aronov, and Zyuzin \cite{AAZ84}. 

We now turn to one--dimensional contacts. 
Then the result (\ref{c_omega_c0})
for $Y(\omega)$ remains valid provided the bare interactions 
$u(q)$ and $v(q)$ in Eq.~(\ref{bare_coulomb_propagator}) are replaced by
$u(q) = 2 e^{2} \ln(1/qa)$ and 
$v(q) = 2 e^{2} \ln(1/q \Delta)$. Further the momentum sum is
one--dimensional, and $C_{0}=[4 \ln(\Delta/a)]^{-1}$
becomes the capacitance per unit length. The zero temperature
differential conductance now takes the form
%
\begin{eqnarray}\label{general_1d}
 \frac{G(V)}{G_{0}} = 1-\mbox{erf}\sqrt{\frac{V_{0}}{V}}
\end{eqnarray}
with 
$V_{0} = \frac{e^{3}}{8 \pi D^{*} C_{0}^{2}} 
[\sum_{j} \lambda_{j}/(1+\xi_{j}^{1/2})]^{2}$.
This result is formally equivalent 
to the conductance of an ultrasmall junction biased via a
RC transmission line \cite{IngoldNato92in} and is in quantitative 
agreement with a
recent experimental study of long tunnel junctions \cite{Pierre01}.

If one of the electrodes is bulky, we again may use 
Eq.~(\ref{dos_independent_general}) to obtain an effective
tunneling density of states of a one--dimensional electrode
%
\begin{eqnarray}\label{tdos_1d}
 \nu(\epsilon) = \nu_{0} \left(
  1-\mbox{erf}\sqrt{\frac{e V_{0}}{\epsilon}} \right) \; .
\end{eqnarray}
While for large energies we recover the perturbative result
$\nu(\epsilon) = \nu_{0} (1 - 2 \sqrt{e V_{0}/\pi \epsilon})$, 
the non--perturbative 
result (\ref{tdos_1d}) does not diverge at low energies but 
approaches an exponential suppression of the densities of states
near the Fermi surface
$\nu(\epsilon) = \nu_{0} \sqrt{\epsilon/\pi e V_{0}} 
\exp(-e V_{0}/\epsilon)$.

For the realistic case of a system with finite size 
the diffusive spreading of the
transfered charge reaches the boundaries of the electrode for 
long times, and then the charge relaxes via the external circuit 
characterized by an impedance $Z_{\rm ext}(\omega)$. 
This modifies the ZBA at very low voltages.
For simplicity we assume that the external charge relaxation, 
which is the only process relevant for 
ultrasmall junctions \cite{Nato92}, is slow
and couples only to the $q=0$ component of the field 
\cite{Kamenev96}. Then Eq.~(\ref{c_omega_c0}) is modified to read
%
\begin{eqnarray}\label{c_omega_finite_size}
 Y(\omega) &=& \frac{e^{2}}{C_{0}} 
  \sum_{q \ne 0} \sum_{j=1}^{2} \frac{\lambda_{j}}
  {(D^{*} q^{2} - i \omega) (D_{j} q^{2} - i \omega)} \nonumber \\
 && \qquad + i \frac{e^{2}}{\omega} \,
  \frac{1}{Z^{-1}_{\rm ext}(\omega) - i \omega C} \; .
\end{eqnarray}
For one--dimensional electrodes of length $L$ the discrete $q$ 
values are of the form $(2 \pi/L) n$, $n$ integer, and 
$C = C_{0}L$ is the total capacitance of the contact, while for
two--dimensional quadratic electrodes $C=C_{0} L^{2}$.

For three--dimensional electrodes of finite size there
are several energy scales in $Y(\omega)$: The charging energy 
$E_{C} = e^{2}/2C$ and for each spatial dimension $\ell$ of the 
electrodes a Thouless energy $E_{\rm Th} = D/\ell^{2}$ as well as a
field Thouless energy $E^{*}= D^{*}/\ell^{2}$. 
Here $\ell$ stands for the length $L$, width $w$, and thickness $a$,
respectively. 
The effective dimension of the electrodes is determined by $E^{*}$. 
Whenever $eV$ exceeds one of the scales $E^{*}_{\ell}$ 
($\ell=L,w,a$) the effective dimension increases by one. 

As an example we consider a symmetric one--dimensional contact 
of length $L$ with an ohmic external impedance $Z_{\rm ext} = R$.
Then, from Eq.~(\ref{c_omega_finite_size}) we obtain 
%
\begin{eqnarray}\label{finite_size_1D_example}
 {\rm Im} \, Y(\omega) 
 &=& {\rm Im} \, Y_{\infty}(\omega) 
  f\left(\frac{\omega}{E^{*}}\right)
  + \frac{\alpha}{1+\alpha^{2} (\omega R_{\rm K} C)^{2}}
\end{eqnarray}
where 
${\rm Im} \, Y_{\infty}(\omega) = e^{2}/[2 \sqrt{2 D^{*}} C_{0}
(1+\xi^{1/2}) \omega^{3/2}]$ 
describes an infinitely long contact, and the function
$f(z)=(g(z)-\xi^{1/2}g(z/\xi))/(1-\xi^{1/2})$ with 
$g(z) = {\rm Im} \, 
  \left[-\sqrt{2 i} \cot\left(\sqrt{i z}/2 \right) \right]$
incorporates all size effects. The last term in 
Eq.~(\ref{finite_size_1D_example}) describes the 
environmental impedance characterized by the dimensionless
parameter $\alpha=R/R_{\rm K}$ with $R_{\rm K}=h/e^{2}$.

\begin{figure}[btp]
\begin{center}
\leavevmode
\epsfxsize=0.45 \textwidth
\epsfbox{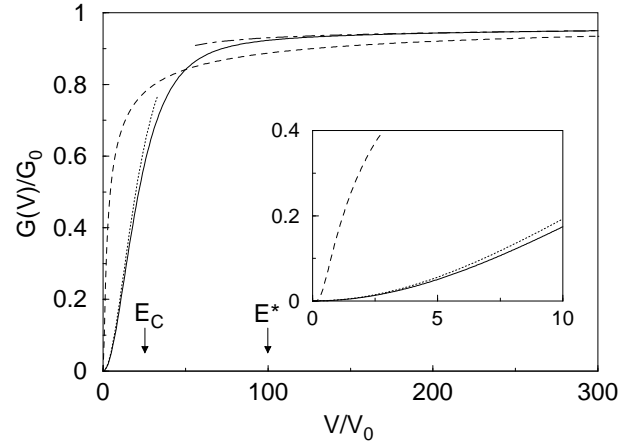}
\vspace*{-.0cm}
\caption{Differential conductance $G(V)$ of a quasi 
one--dimensional tunnel junction of finite length with an 
external impedance at $T=0$ (solid line), 
see text for details. 
This is compared with the results of a 
zero--dimensional (dotted line) and a long one--dimensional
junction (dashed line). The chain dotted line corresponds
to the perturbative result.}
\label{gv_figure}
\end{center}
\end{figure}

In Fig.~\ref{gv_figure} we present results for typical values 
of metallic quasi one--dimensional tunnel junctions
\cite{Pierre01} 
($\xi=4 \times 10^{-4}$, $E^{*}/eV_{0} = 100$, $E_{C}/eV_{0}=25$).
The full line corresponds to a finite length junction 
shunted by a rather large environmental impedance ($\alpha=1$)
to display the crossover between zero--dimensional and 
one--dimensional behavior clearly. The dotted line gives 
the conventional environmental Coulomb blockade of a 
zero--dimensional junction which determines the behavior 
for voltages below $E_{C}/e$ (see inset). 
On the other hand, for large 
voltages above $E^{*}/e$ the result (\ref{general_1d}) for a 
long one--dimensional junction (dashed line) is approached.
The chain dotted line results from a perturbative calculation 
of the diffusive anomaly \cite{AA79}.

In summary, we have studied tunneling in large junctions with
diffusive electron motion. Treating the Coulomb interaction
non--perturbatively, the $I$--$V$--curve was written
in a form familiar from Coulomb blockade theory for ultrasmall
junctions. Interaction effects are again incorporated in a spectral
function $P(\epsilon)$.
However, while for ultrasmall
junctions $P(\epsilon)$ depends only on the effective impedance
$[Z^{-1}_{\rm ext}(\omega) - i \omega C]^{-1}$, for large junctions 
this quantity is replaced by $-i \omega Y(\omega)/e^{2}$ which 
incorporates  also the diffusive anomaly of disordered metals.

We would like to thank the authors of Ref.~\cite{Pierre01} and 
G. G\"oppert for valuable discussions. One of us (JR) is grateful to
the CEA-Saclay for hospitality during a stay where parts of this work
were carried out. Financial support was provided by the DAAD. 

\begin{thebibliography}{12}
\bibitem{remark_spin}
We restrict ourselves to spinless fermions here but include spin 
degeneracy by appropriate factors of 2.
\bibitem{remark_delta}
$\Delta$ is actually the dipole moment 
per unit charge which due to the finite film thickness  
is somewhat larger than the actual barrier thickness.
\end{thebibliography}
\end{document}